\documentclass[aps,11pt]{revtex4}
\usepackage[dvips]{graphicx}
\begin{document}

\title{Violation of the Curie law in Na$_{2}$V$_{3}$O$_{7}$ \\
as the crystal-field and spin-orbit coupling effect$^{+}$}
\author{R. J. Radwanski}
\homepage{http://www.css-physics.edu.pl}
\email{sfradwan@cyf-kr.edu.pl}
\affiliation{Center of Solid State Physics, S$^{nt}$Filip 5, 31-150 Krakow, Poland,\\
\& Institute of Physics, Pedagogical University, 30-084 Krakow,
Poland}
\author{Z. Ropka}
\affiliation{Center for Solid State Physics, S$^{nt}$Filip 5, 31-150 Krakow, Poland}
\begin{abstract}
We have shown that the observed drastic violation of the
Curie-Weiss law in Na$_{2}$V$_{3}$O$_{7}$, reported in Phys. Rev.
Lett. 90 (2003) 167202, is caused by conventional crystal-field
interactions and the intra-atomic spin-orbit coupling of the
V$^{4+}$ ion. The fine discrete electronic structure of the
3$d^{1}$ configuration, with a substantial orbital moment, is the
reason for anomalous low-temperature properties of
Na$_{2}$V$_{3}$O$_{7}$. According to the Quantum Atomistic
Solid-State Theory (QUASST) Na$_{2}$V$_{3}$O$_{7}$ is expected to
exhibit pronounced heavy-fermion phenomena at low temperatures.
This study confirm our earlier claim that the orbital moment has
to be unquenched in description of 3$d$-atom compounds.



\end{abstract}
\maketitle

Temperature dependence of the paramagnetic susceptibility $\chi (T)$ of Na$_{2}$V$_{3}$O$_{7}$ exhibits at temperatures
below 100 K a drastic violation of the Curie-Weiss law \cite{1}. From the experimentally measured temperature
dependence of the magnetic susceptibility Gavilano \textit{et al.}
have inferred that the effective moment of the V$^{4+}$ ion in Na$_{2}$V$%
_{3}$O$_{7}$ is reduced by the one order of magnitude upon reducing the temperature from 100 to 10 K. In the figure 1
of Ref. \cite{1} it is seen that after taking into account the diamagnetic contribution $\chi _{o}$ the inverse
susceptibility shows in the temperature range 100-300 K a straight line behavior with the effective moment p$_{eff}$ of
1.9 $\mu _{B}$ per V ion. Another straight line between 20 and 1.9 K implies $p_{eff}$ of one order of magnitude
smaller. Gavilano \textit{et al.} provide an explanation that ''The reduction of the effective magnetic moment is most
likely due to a gradual process of moment compensation via the formation of singlet spin configurations with most but
not all of the ions taking part in this process. This may be the result of antiferromagnetic interactions and
geometrical frustration.'' They further conjectured ''the compensation of eight out of the nine V spins ...'' in order
to reproduce the observed reduction of the effective moment by one order of magnitude. They also found that
Na$_{2}$V$_{3}$O$_{7}$ shows no sign of the magnetic order down to 1.9 K - we find this experimental observation to be
in sharp contradiction with the earlier Gavilano \textit{et al.}'s conclusion about the presence of strong
antiferromagnetic interactions.

The aim of this Letter is to propose a more physical explanation for this abnormal temperature behavior of the
paramagnetic susceptibility of Na$_{2}$V$_{3}$O$_{7}$. Namely we argue that this drastic violation of the Curie(-Weiss)
law can be understood as caused by well-known conventional phenomena like the crystal field (CEF)\ interactions but
with taking into account the intra-atomic spin-orbit (s-o) coupling. Calculations of the electronic structure
associated with the V$^{4+}$ ion (3$d^{1}$ configuration) under the action of the CEF\ and s-o interactions reveal that
the ground state can have, for instance, quite small value of the magnetic moment - it results in the drastic departure
of $\chi (T)$ from the Curie law. It is in agreement with our earlier finding that even weak s-o coupling unquench a
quite large orbital moment \cite{2,3}. Here we present results of our studies on influence of the spin-orbit coupling
and the off-octahedral trigonal distortions on the temperature dependence of the paramagnetic susceptibility. These
studies reveal the drastic violation of the Curie law. Our results can be quite obvious for some people, in particular
for those having experience with CEF effects in rare-earth compounds and knowing a 70-year-old book of Van Vleck
\cite{4}, but this paper is motivated by the recent paper by Gavilano \textit{et al.} in Phys. Rev. Lett. \cite{1},
where very exotic explanation for the anomalous temperature dependence of the paramagnetic susceptibility of
Na$_{2}$V$_{3}$O$_{7}$ has been published. The very direct motivation for this paper is the subsequent rejection by the
Editor of Phys. Rev. Lett. of our paper ''Spin-orbit origin of large reduction of the effective moment in
Na$_{2}$V$_{3}$O$_{7}$'' \cite{5} who found the proposed CEF+s-o explanation for the violation of the Curie law in
$\chi (T)$ so unusual that it does not deserve for sharing with the scientific community. Rejection arguments of the
Editor and referees of Phys. Rev. Lett. (the correspondence is attached here as Appendix) show that very obvious and
natural for us, and we thought that also for others, theoretical approach, pointing out the importance of the crystal
field and the spin-orbit coupling on the temperature dependence of the paramagnetic susceptibility, is not accepted by
one of the most prestigious physical journal. It shows that behind the rejection there is very serious scientific
problem about the role played by the spin-orbit interactions in 3$d$-atom containing compounds. In fact, this problem
started in Phys. Rev. Lett. already in 1997 when we have submitted our first paper on the relativistic spin-orbit
effect on the electronic structure of 3$d$ paramagnetic ions \cite{6}. The rejection of the normal submission of Ref.
\cite{5} has forced us to write 3 November 2003 the Comment on "Low-dimensional spin $S$=1/2 system at the quantum
critical limit: Na$_{2}$V$_{3}$O$_{7}$" \cite{7}.

Coming to Na$_{2}$V$_{3}$O$_{7}$ there is the general agreement about the existence of the V$^{4+}$ ions in
Na$_{2}$V$_{3}$O$_{7}$ but according to our approach we described the one 3$d$ electron in the V$^{4+}$ ion by quantum
numbers $L$=2 and $S$=1/2 that are coupled by the intra-atomic spin-orbit coupling \cite{8}. The 10-fold degeneracy of
the charge-formed ground term $^{2}D$ is removed by the intra-atomic spin-orbit interactions $H_{s-o}$ and in a solid
by crystal-field interactions $H_{CF}$. This situation can be exactly traced by the consideration of a single-ion-like
Hamiltonian
\begin{equation}
 H_{d}=H_{CF}+H_{s-o}+H_{Z}=\sum B_{n}^{m}O_{n}^{m}+\lambda L\cdot S+\mu _{B}(L+g_{e}S)\cdot \mathbf{B}\,\;\;\;
\end{equation}
in the 10-fold degenerated spin-orbital space provided the symmetry of the local surroundings is known. Although we
knew that the local symmetry in Na$ _{2}$V$_{3}$O$_{7}$ is quite complex we approximate, for simplicity in this paper,
the CEF\ interactions at the V site by considering dominant octahedral interactions with a trigonal distortion. Then
the Hamiltonian takes a form:
\begin{equation}
\label{eq:ham}
H_{d}=H_{CF}^{octa}+H_{s-o}+H_{CF}^{tr}+H_{Z}=-2/3B_{4}^z(O_{4}^{0}-20\sqrt{2}O_{4}^{3})+\lambda L\cdot
S+B_{2}^{0}O_{2}^{0}+\mu _{B}(L+g_{e}S)\cdot \mathbf{B} \,\;\;\;
\end{equation}
For the octahedral crystal field we take $B_{4}^z$= +200 K in the Hamiltonian with z-axis taken along the cube edge.
The sign ''+'' in $B_{4}^z$ comes up from \textit{ab initio} calculations for the ligand octahedron \cite{6}. The
spin-orbit coupling parameter $\lambda _{s-o}$ is taken as +360 K, as in the free V$^{4+}$ ion \cite{9}.

The resulting electronic structure of the 3$d^{1}$ ion contains 5 Kramers doublets separated in case of the dominant
octahedral CEF interactions into 3 lower doublets, the $T_{2g}$ cubic subterm, and 2 doublets, the $E_{g}$ subterm,
about 2 eV above (Fig. 1). The $T_{2g}$ subterm in the presence of the spin-orbit coupling is further split into a
lower quartet and an excited doublet, Fig. 1(2). The lower quartet is split by off-octahedral distortions into two
Kramers doublets. It is important to realize that whatever lower symmetry is only 5 Kramers doublets always are for the
$d^{1}$ configuration. Positive values of the trigonal distortion parameter $B_{2}^{0}$ yields the ground state that
has a small magnetic moment, Fig. 1(3). For $B_{2}^{0}$ = +9 K the ground state moment amounts to $\pm 0.21$ $\mu
_{B}$. It is composed from the spin moment of $\pm 0.46$ $\mu _{B}$ and the\ orbital moment of $\mp 0.25$ $\mu _{B}$
(antiparallel). The sign $\pm $ corresponds to 2 Kramers conjugate states. The excited Kramers doublet lies at 58 K (5
meV) and is almost non-magnetic - its moment amounts to $\pm 0.03$ $\mu _{B}$ only (=$\pm 1.03$ $\mu _{B}+2\cdot (\mp
0.50$ $\mu _{B})$) due to the cancellation of the spin moment by the orbital moment. So small and so different moments
for the subsequent energy levels is the effect of the spin-orbit coupling and local distortions.\\
The paramagnetic susceptibility is calculated by the definition as $\chi (T)$ $=$ $-\partial ^{2}F(T,B)/\partial
B^{2}$, where $F(T,B)$ is the Helmholtz free energy calculated from the discrete electronic structure resulting from
Hamiltonian (\ref{eq:ham}).
\begin{figure}[!ht]
\includegraphics[height = 11.5 cm]{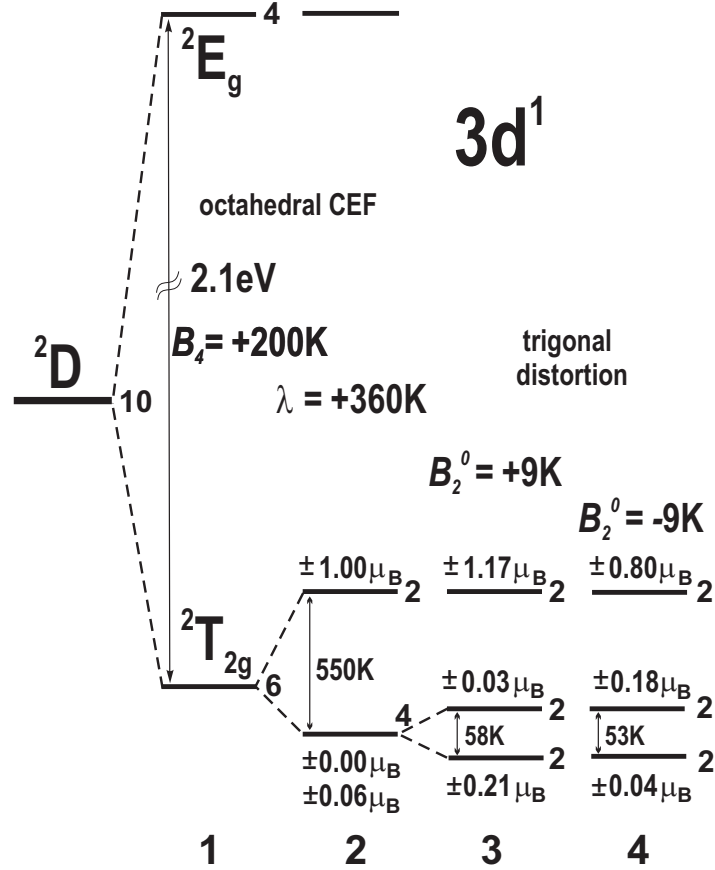}
\caption{Calculated localized states of the 3$d^{1}$ configuration in the V$^{4+}$ ion under the action of the crystal
field and spin-orbit interactions originated from the 10-fold degenerated $^{2}D$ term; (1) the splitting of the
$^{2}D$ term by the octahedral CEF\ surroundings with $B_{4}$=+200 K, $ \protect\lambda _{s-o}$ =0; (2) the splitting
by the combined octahedral CEF and spin-orbit interactions; (3) and (4) show the effect of the trigonal distortion
$B_{2}^{0}$=+9 K (3) and $B_{2}^{0}$=-9 K (4). The states are labelled by the degeneracy in the spin-orbital space and
the value of the magnetic moment. Scheme (3) provides exceptionally good fit to the experimental $\chi (T)$ of
Na$_{2}$V$_{3}$O$_{7}$.}
\end{figure}
\begin{figure}[ht]
\includegraphics[width = 0.92\textwidth]{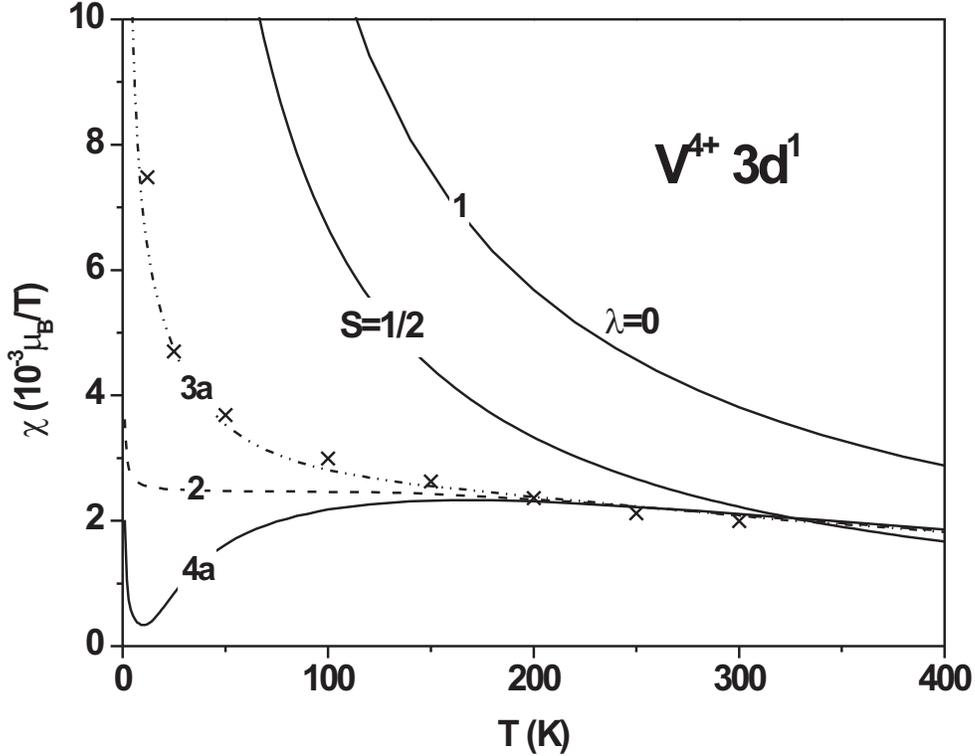}
\caption{The calculated temperature dependence of the atomic-scale paramagnetic susceptibility $\protect\chi (T)$ for
the 3$d^{1}$ configuration of the V$^{4+}$ ion for different physical situations: \ line
(1) - $\protect\chi (T)$ for the purely octahedral crystal field with $B_{4}$%
=+200 K without the spin-orbit coupling ($\protect\lambda _{s-o}$= 0); curve (2) - the octahedral crystal field in
combination with the spin-orbit coupling $\protect\lambda _{s-o}$= +360 K; curves (3a) and (4a) show the influence of
the off-octahedral trigonal distortions $B_{2}^{0}$=+9 K (3) and $B_{2}^{0}$= -9 K (4) on $\protect\chi (T)$ parallel
to the cube diagonal; the curve (3a) reproduces very well measured experimental data of $\chi (T)$ of
Na$_{2}$V$_{3}$O$_{7}$ (x), after Refs \protect\cite{1,10}, with taking into account the diamagnetic term $\protect\chi
_{o}$ of -0.0007 $\protect\mu _{B}$/T V-ion ($\simeq $ -0.0004 emu/mol V) and the number of V atoms (for molar units it
is the Avogadro number).}
\end{figure}
\begin{figure}[ht]
\includegraphics[width = 0.92\textwidth]{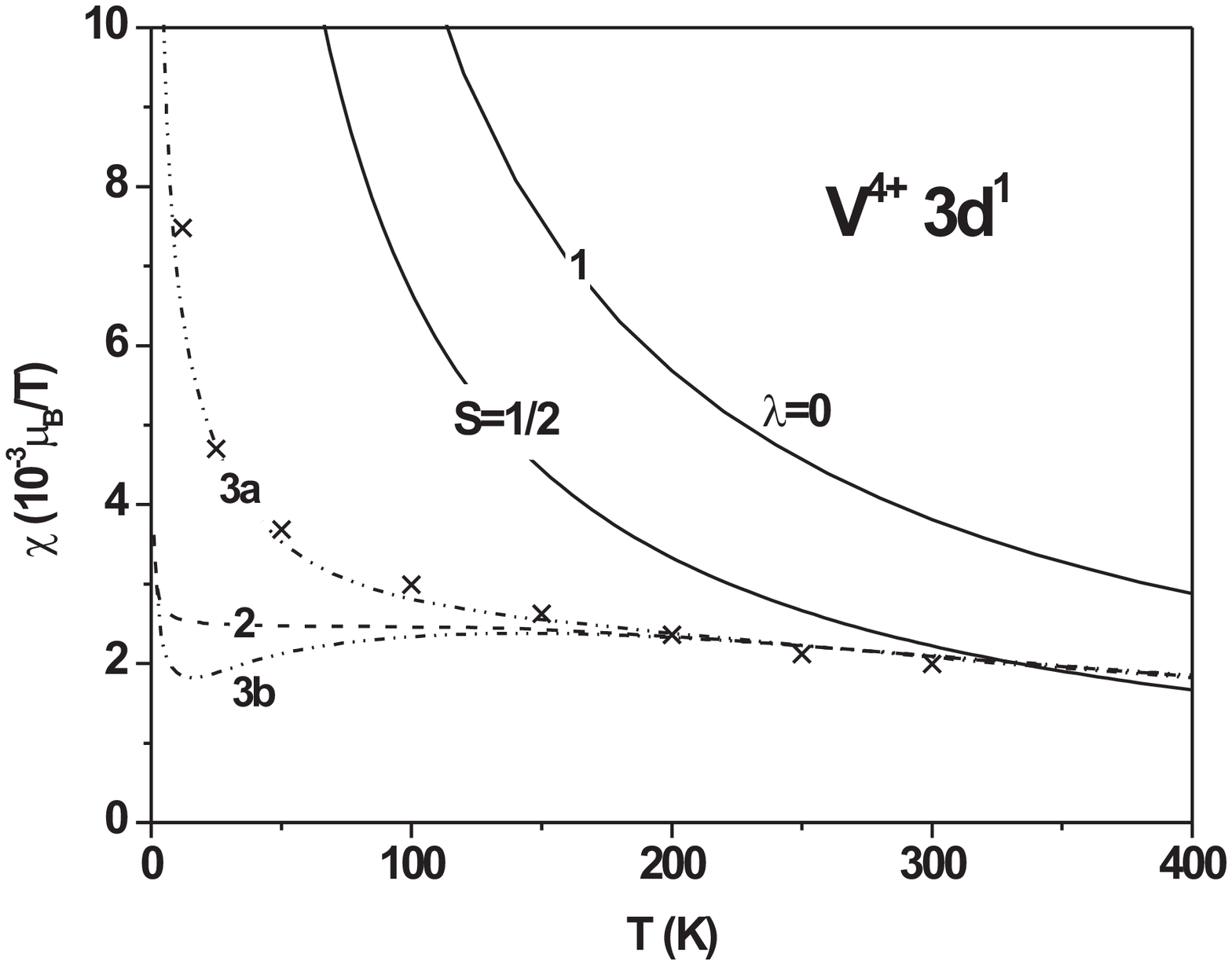}
\caption{The calculated anisotropy of the temperature dependence
of the atomic-scale paramagnetic susceptibility for the 3$d^{1}$
configuration of the V$^{4+}$ ion for magnetic field parallel (3a)
and perpendicular (3b) to the trigonal distortion axis for
$B_{4}$=+200 K, $ \protect\lambda _{s-o}$ =360 K and trigonal
$B_{2}^{0}$=+9 K. Description of other curves is like in Fig. 2.}
\end{figure}
In Fig. 2 the calculated results for the paramagnetic susceptibility are shown for different physical situations of the
V$^{4+}$ ion: line (1) - for the purely octahedral crystal field without the spin-orbit coupling (p$_{eff}$ = 2.29 $\mu
_{B}$); a value of p$_{eff}$ = 3.0 $\mu _{B}$ is found for the free $d^{1}$ configuration; curve (2) shows $\chi (T)$
in the octahedral crystal field in the presence of the spin-orbit coupling; curves (3) and (4) - illustrate the effect
of trigonal off-octahedral distortions: curve (3) for $B_{2}^{0}>0$ (=+9 K), and curve (4) for $B_{2}^{0}<0$ (= -9 K).
Everybody can see that all of these curves are completely different from the Curie-law $S$=1/2 behavior, expected for
the collection of the free (i.e. non-interacting magnetically) one-electron spins. It means that the customary
treatment of the V$^{4+}$ ion with one 3$d$ electron as the $S$ =1/2 system, i.e. with the spin-only magnetism and with
taking into account the spin degree of freedom only, is completely not justified. It is simply wrong. The neglect in
the current literature of the orbital moment is consistent with the widely-spread conviction that the orbital magnetism
plays rather negligible role due to the quenching of the orbital moment for 3$d$ ions. Na$_{2}$V$_{3}$O$_{7}$ is an
example of numerous compounds in which the $S$=1/2 behavior in the temperature dependence of the paramagnetic
susceptibility is drastically violated (CaV$_{4}$O$_{9}$, MgVO$_{3}$, (VO)$_{2}$P$_{2}$O$_{7}$, ...), in particular at
low temperatures. Our calculations prove that even weak spin-orbit coupling and small distortions of the local
surroundings of the V$^{4+}$ ion cause very drastic change of $\chi (T)$ at low temperatures in comparison to the Curie
law. This violation of Curie law is the best seen in the change of the slope of the $\chi ^{-1}(T)$ plot below 100 K,
see Fig. 3 of Ref. \cite{5}, curves (2) and (5). The curve (3a), with $B_{4}$ =+200 K, the spin-orbit coupling
$\protect\lambda _{s-o}$= +360 K and $B_{2}^{0}$ =+9 K, reproduces very well the measured experimental data of $\chi
(T)$ of Na$_{2}$V$_{3}$O$_{7}$(x) \cite{1,10}. Basing on these results we are convinced that the strong temperature
dependence of the effective moment, inferred in Ref. \cite{1} and attributed to the special nanotube structure realized
in Na$_{2}$V$_{3}$O$_{7}$, predominantly results from the intra-atomic spin-orbit coupling and local distortions.

In Fig. 3 the calculated anisotropy of $\chi (T)$ is shown by presenting the calculated paramagnetic susceptibility for
the magnetic field applied parallel (denoted by "3a") and perpendicular (denoted by "3b") to the distortion direction.
For a polycrystalline material the measured susceptibility will follow that given by the larger susceptibility
direction, i.e. along the curve 3a in the present case. The obtained agreement with the experimental data is quite
remarkable. We treat this coincidence as not fully relevant owing to the much more complex local symmetry of the V$
^{4+}$ ion in Na$_{2}$V$_{3}$O$_{7}$, to a large uncertainty in the evaluation of the diamagnetic term and of the
paramagnetic susceptibility measured on a polycrystalline sample. We take, however, the reached agreement as strong
argument for the high physical adequacy of the applied QUASST approach \cite {11,12,13} and as strong indication for
the existence of the fine electronic structure in Na$_{2}$V$ _{3}$O$_{7}$, originating from the V$^{4+}$ ion,
determined by crystal-field and spin-orbit interactions. Very important is the fact that our approach is able to
reproduce the overall $\chi (T)$ dependence in the full measured temperature range and that it reproduces the absolute
value of the macroscopic magnetic susceptibility. For the recalculation of the microscopic atomic-scale susceptibility
we take into account only the number of the V ions involved - here for the molar susceptibility the simply Avogadro
number. The used parameters $B_{4}$ (+200 K), $\lambda _{s-o}$ (+360 K) and $B_{2}^{0}$ (+9 K) have clear physical
meaning. The value of +200 K for $B_{4}$ is consistent with the value of +260 K found recently experimentally for the
octahedral crystal field in LaCoO$_{3}$ \cite{14} - the similarity is related to the fact that the strength of $B_{4}$
is predominantly determined in both cases by the local oxygen octahedron. The obtained good reproduction of such
nontrivial experimental results we take as further indication for the high physical adequacy of the QUASST\ conjecture
that the electronic and magnetic properties of the 3\textit{d-}ion containing compounds are predominantly determined by
the fine electronic structure of the 3\textit{d} ion in the meV scale. According to the QUASST theory the atomic-like
electronic structure is preserved in 3$d$-atom containing compounds, where the 3$d$ ion is the full part of the
crystal. In Ref. \cite{5} it was pointed out that QUASST predicts Na$_{2}$V$_{3}$O$_{7}$ to exhibit heavy-fermion-like
properties in the specific heat at ultra-low temperatures. They are related to the removal of the Kramers spin-like
degeneracy of the ground state, not removed down to 1.9 K due to the lack of the magnetic order.

In conclusion, we argue that the experimentally-observed temperature dependence of the paramagnetic susceptibility of
Na$_{2}$V$_{3}$O$_{7}$ with strong violation of the Curie law at low temperatures can be remarkably well explained
within the single-ion approach, extended by us to the Quantum Atomistic Solid-State Theory QUASST for 3$d$-atom
containing compounds, taking only into account conventional local atomic-scale effects like the crystal-field
interactions, the intra-atomic spin-orbit coupling and the orbital 3$d$ magnetism. The used parameters $B_{4}$ (+200
K), $\lambda _{s-o}$ (+360 K) and $B_{2}^{0}$ (+9 K) have clear physical meaning. The superiority of our explanation
relies in the fact that it explains consistently both zero-temperature properties as well as thermodynamics and, the
most important, it is based on well-known physical concepts. \\
$^{+}$ This paper is an effect of the rejection 16-10-2003 by Editor of Phys.Rev.Lett. (PRL) of our paper
\textit{Spin-orbit origin of large reduction of the effective moment in Na$_{2}$V$_{3}$O$_{7}$}
(http://arXiv/cond-mat/0309460). The original paper \textit{To the origin of large reduction of the effective moment in
Na$_{2}$V$_{3}$O$_{7}$} can be found in http://arXiv/cond-mat/0307272. In cond-mat/0309460, as APPENDIX A-F, one can
found the full documentation of the Phys.Rev.Lett. procedure as well as two referee reports and our answers. Here the
subsequent correspondence is attached. About the abnormal situation we have informed the Editor-in-Chief and the
President of the American Physical Society. The restriction of the scientific discussion on just published papers we
treat as very serious violation of the scientific rules. We would like to inform that we have agreed for publishing of
our paper together with the negative referee reports. We remind, we chose Phys. Rev. Lett. because the paper of
Gavilano \textit{et al}., reporting properties of Na$_{2}$V$_{3}$O$_{7}$ with an exotic explanation, has appeared in
Phys. Rev. Lett. in April 2003 (\textbf{90}, 167202).

In the situation of the rejection of the original papers, on basis
of nonphysical and wrong scientific arguments we submit 31 October
2003 to Phys. Rev. Lett. the Comment to Ref. 1 - this Comment is
available in cond-mat/0311033. This Comment was recently, 17
November 2003, also rejected. It is the enormous violation of the
scientific rules - the publication of Comment is the obligation of
each journal, which regards itself to be the scientific journal.
In such the abnormal situation we decided to print all
correspondence with Phys. Rev. Lett. (also available at
www.css-physics.edu.pl) - then all physicists can judge scientific
arguments of both sides and realize the clue of our controversy
with Phys. Rev. Lett.. We maintain our scientific view that for
the description of 3d-/4f-/5f-atom containing compounds the local
electronic structure, determined by strong on-site electron
correlations, the intra-atomic spin-orbit coupling, crystal-field
interactions and local surroundings, have to be considered at the
beginning of each analysis of physical properties of the whole
compound. We named this idea as Quantum Atomistic Solid State
Theory (QUASST \cite {11,12,13}). We are surprised that such
obvious physical concept, basing on the atomistic theory of
matter, and the used scientific methodology, the well-known
phenomena analyzed the first, is discriminated in Phys. Rev.
Lett., one of the most prestigious physical journal. It seems to
be the next case in the history of Science that obvious idea has
been rejected by the leading scientists. We have to add that for
us this idea was obvious and well-established in physics and
chemistry, let mention works of Bethe, Kramers, Van Vleck and
many, many others working on the crystal field from 1929. We
undertook these calculations for 3d-atom compounds in year of 1996
with the modest aim to perform exact calculations of the influence
of the spin-orbit coupling on the low-energy states of the 3$d$
paramagnetic ions (Phys. Rev. B BZR586; Phys. Rev. Lett. LA6567;
LE6925 - Ref. \cite{5}), which have been done up to that time
badly by the perturbation method only. The rejection politics of
the Editors of Phys. Rev. and the rejection arguments of the
referees of Phys. Rev., that must be the best present physicists,
reveal that this atomistic concept is far from being presently
accepted. Indeed, everybody can see that the atomistic concept has
been somewhere lost in nowadays theories of the solid-state
physics and chemistry. We point out that we do not insist on the
Editor and on our colleague scientists to accept our ideas and our
methodology, but we insist that our scientific point of view
should get the standard scientific treatment and cannot be
discriminated, freely prohibited and rejected in scientific
journals. Why so brilliant physicists are not able to openly prove
errors in our approach? Than it will be the end of the
Radwanski\&Ropka story, created by the Phys. Rev.'s Editors. We
are ready to accept real scientific arguments.

Added: 17 November 2004: The above version was prepared a year
ago: it was put to ArXiv 25-11-03 as 0311575 and later 0401127,
0401153 and 0401196 (full version is available on our homepage).
We put it again now as in the current literature the electronic
structure for the 3d$^{1}$ configuration in YTiO$_{3}$ and
LaTiO$_{3}$ or in other V$^{4+}$ ion systems is still under
discussion (Phys.Rev. B 69 (2004) 134403, for instance). With
satisfaction we note that the importance of the spin-orbit
coupling and the detailed local surroundings (crystal field) start
to be also noticed at present in Phys.Rev. B and Phys.Rev.Lett..


\begin{thebibliography}{99}
\bibitem{8} the use of $L$ and $S$ or $l$ and $s$ does not make difference for the  $d^{1}$ configuration, but the important
difference occurs for other configurations with $n\geq2$.
\end{thebibliography}
\end{document}